\documentclass[aip,pof,twocolumn,reprint,amsmath,amssymb]{revtex4-1}
\usepackage{graphicx}
\usepackage{dcolumn}
\usepackage{bm}
\usepackage{hyperref}
\usepackage{color}

\begin{document}

\preprint{AIP/123-QED}

\title[Electro/Magnetically Induced Rotation]{Electro/Magnetically Induced Controllable Rotation In Small-scale Liquid Flow}

\author{A. Amjadi}
 \affiliation{Physics Department, Sharif University of Technology, Tehran, Iran}
\author{A. Nejati}
\affiliation{Physikalisches Institut and Bethe Center for Theoretical Physics, Universit\"{a}t Bonn, Germany}
\author{S. O. Sobhani}
\email{omid.sobhani@gmail.com}
\affiliation{Physics Department, Sharif University of Technology, Tehran, Iran}
\author{R. Shirsavar}
\affiliation{Department of Physics, Faculty of Science, University of Zanjan, Zanjan, Iran}

\date{\today}

\begin{abstract}
We study all the possibilities of producing rotating flow in an incompressible fluid by electric and magnetic fields. We start with a general theoretical basis and look for different configurations and set-ups which electric/magnetic field and an electric current affect the vorticity of fluid resulting in rotation on liquid flow. We assume steady-state conditions and time-independent electric and magnetic fields as the external body torque. Regarding the theoretical basis, we propose three experimental set-ups in which by applying fields on a fluid, rotational vortices are produced: (a) a uniform electric field and a uniform electric current, (b) a uniform electric current and a non-uniform magnetic field, and (c) a non-uniform electric current and a uniform magnetic field. The first case has been reported in detail named "Liquid Film Motor". The two other cases are experimentally investigated here for a cubic an cylindrical cells. The rotational velocity patterns are obtained by PIV technique, and the results are discussed and justified by a preliminary estimation based on the torque exerted by magnetic fields on electric currents.
From the log-log plot of angular velocity versus current, the non-linearity factors of the rotational flow for cylindrical and rectangular geometries are obtained. 
\end{abstract}


\maketitle

\section{Introduction}
The prospect of manipulating the fluid flow by electric and magnetic fields with its various applications such as washers, mixers and drug delivery systems, have been the motivation for a wide range of studies on the electrically- and magnetically-driven fluid flow. Interaction of electric fields with fluids have been extensively studied by Melcher \emph{et. al.} ~\cite{melcher1961,melcher1969,washabauch1989}, and electro-hydrodynamical instabilities has been investigated by Taylor ~\cite{taylor1966} and Saville \cite{saville1997}. Related experimental studies also have been performed on electric fields to control and manipulate nano- and micro-sized particles in a fluid by Ramos \emph{et. al.} ~\cite{ramos1999ac,ramos1999ab}.Faetti \emph{et. al.}~\cite{faetti1983,faetti1979} have investigated vortices which are produced on a suspended liquid crystal film while a sufficiently large electric current passes through the film. Morris \emph{et. al.}~\cite{morris1990,daya1997} have suggested that electric charges accumulated on the surface of suspended films produce these instabilities. Amjadi \emph{et. al.}~\cite{amjadi2009,shirsavar2011,shirsavar2012} have proposed a method for controlling the rotational flow on suspended films of fluid called "Liquid Film Motor". They have demonstrated that a simultaneous application of an electric field and an electric current produces a controllable rotation in suspended films. Liu \emph{et. al.}~\cite{liu2012,liu2013} have theoretically investigated such instabilities in presence of alternating fields and studied vibration in these systems.
Fluid flow behaviour of ferrofluids is also investigated in presence of magnetic fields by Zahn \emph{et. al.}~\cite{zahn2001} and Chaves \emph{et. al.}~\cite{chaves2006}.

In this paper, we investigate further all the possible cases in which electric/magnetic fields and electric current on the fluid could produce rotation. In the next section, we provide a theoretical background for the experimental set-ups to show the effect of external electric/magnetic fields and currents on the vorticity of a fluid. The set-ups are described in section~\ref{sec:expapparat}.
Section~\ref{sec:results} describes our observations that when an electric or magnetic field is applied on a fluid carrying an electric current, a rotational flow is produced. These methods are applicable to surface and bulk liquids in macroscopic scales.
As it is expected the direction and velocity of the flow can be controlled by the direction and strength of the electric/magnetic fields and the electric current. The results are justified by a preliminary estimation based on the force exerted by electric/magnetic fields on electric currents, $\rho \vec{E} + \vec{J} \times \vec{B}$.

The experimental results can provide a basis for novel techniques to design dynamic micro-pumps that are based on electrohydrodynamic and magnetohydrodynamic effects. Such devices have various applications such as micro- mixing, washing, pumping, especially in biology~\cite{amirouche2009}, \emph{e.g.} for drug delivery devices~\cite{nisar2008}, or more generally, in the fast-growing `lap-on-chip' technology~\cite{laser2004}.

\section{Theoretical Basis}
\label{sec:theory}

In some fluid engineering applications, one needs to produce a controlled rotational flow in a fluid. As a local measure of rotation, vorticity is one of the major dynamical properties of a fluid; it is defined as $ \vec{\omega}  = \nabla \times \vec{v}$, where $\vec{v}$ is the velocity field of the fluid. Furthermore, fluid circulation $\gamma$, the macroscopic measure of rotation is the surface integral of vorticity, \emph{i.e.} $\gamma = \oint_C ~ \vec{v} \cdot \text{d}\vec{l} = \iint_S \vec{\omega} \cdot \text{d}\vec{s} $, where $S$ is the surface area inside an arbitrary closed curve $C$. The evolution of vorticity can be derived directly from the Navier-Stokes equation:

\begin{align}
    \frac{\partial \vec{\omega}}{\partial t} + (\vec{v} \cdot \nabla) \vec{\omega} =& (\vec{\omega} \cdot \nabla)\vec{v} - \vec{\omega} (\nabla \cdot \vec{v}) + \frac{1}{\rho^{2}} \nabla \rho \times \nabla P  \nonumber  \\
      & + \nabla \times \left(\frac{\nabla \cdot \tau} {\rho} \right) + \nabla \times \vec{F} ~, \label{1}
\end{align}
where $\rho$ is the density, $P$ is the pressure, $\tau$ is the viscous stress tensor and $\vec{F}$ is the ``body force'' term.

In steady-state conditions ($\frac{\partial \vec{\omega}}{\partial t} = 0$), for an incompressible liquid, $\nabla \cdot \vec{v} = 0$, and since the fluid is assumed to be homogeneous, $\nabla\rho = 0$. Also for an incompressible, homogeneous and Newtonian fluid, $\nabla \cdot \tau = \mu \nabla^{2} \vec{v}$, where $\mu$ is the dynamic viscosity of the fluid. Then Eq.~\ref{1} reduces to:
\begin{equation}
    \nabla \times {\vec F} = ({\vec v} \cdot \nabla) {\vec \omega} - ({\vec \omega} \cdot \nabla) {\vec v} - \nabla \times (\mu \nabla^{2} {\vec v}) ~ . \label{2}
\end{equation}
Eq.~\ref{2} is a relation between $\vec v$, ${\vec \omega}$ and $\vec{F}$, and indicates that a \emph{rotational} force, for which $\nabla \times \vec{F} \neq 0$, would change the vorticity of a fluid. In contrast, non-rotational forces like an electric charge in electric field cannot directly produce such a change.

In our experiments, we deal with electric/magnetic forces $\vec{F}_{E} + \vec{F}_{M}$ which are rotational forces if $\nabla \times(\vec{F}_{E} + \vec{F}_{M})$ is non-vanishing. The electric/magnetic force per unit volume in a fluid can be obtained as
\begin{equation}
\left\{
\begin{array}{lr}
\vec{F}_{E}=\rho_e \vec{E} + \nabla(\vec{P} \cdot \vec{E}) ~, \\
\vec{F}_{M}= \vec{J} \times \vec{B} + \nabla(\vec{m} \cdot \vec{B}) ~,
\end{array} \right.
\label{fEM}
\end{equation}
where $\vec{E}$ and $\vec{B}$ are electric and magnetic fields, $\rho_e$ and $\vec{J}$ are electric charge and current densities, and $\vec{P}$ and $\vec{m}$ are electric and magnetic dipole moments. By applying the curl to Eq.~\ref{fEM}, since $\nabla \times \nabla\phi = 0$, one obtains
\begin{equation}
\left\{
	\begin{array}{lr}
    \nabla \times \vec{F}_{E} = \nabla \times (\rho_e \vec{E}) ~, \\
    \nabla \times \vec{F}_{M} = \nabla \times (\vec{J} \times \vec{B}) ~.
    \end{array}  \right.
  \label{cfEM1}
\end{equation}
more explicitly,
\begin{align}
\left\{
\begin{array}{lr}
\nabla \times(\rho_e \vec{E})=\rho_e \nabla \times \vec{E} + \nabla \rho_e \times \vec{E} ~ \\
    \nabla \times (\vec{J} \times \vec{B})= \big( (\nabla \cdot \vec{B}) + (\vec{B} \cdot \nabla) \big) \vec{J}  \\
     \qquad  - \big( (\nabla \cdot \vec{J}) + (\vec{J} \cdot \nabla) \big)\vec{B} ~.
\end{array}\right.
\end{align}
Then using Maxwell equations,
\begin{equation}
\left\{
	\begin{array}{lr} 
    \nabla\times \vec{E} = -\dfrac{\partial \vec{B}}{\partial t} ~, \\
    \nabla \cdot \vec{B} = 0 ~,
    \end{array} \right.
\end{equation}
along with the charge conservation equation,
\begin{equation}
    \nabla\cdot \vec{J} = -\dfrac{\partial\rho_e}{\partial t}~,\label{chrgcons}
\end{equation}
one obtains:
\begin{equation}
\left\{
\begin{array}{lr}
\nabla \times \vec{F}_{E} = \nabla \rho_e \times \vec{E} + \rho_e \nabla \times \vec{E} ~, \\
   \nabla \times \vec{F}_{M} = (\vec{B} \cdot \nabla) \vec{J} - (\vec{J} \cdot \nabla) \vec{B} + (\dfrac{\partial\rho_e}{\partial t}) \vec{B} - \rho_e(\dfrac{\partial \vec{B}}{\partial t}) ~.
\end{array} \right.
\label{cfEM2}
\end{equation}
For steady-state flow conditions, \emph{i.e.} $\frac{\partial\rho_e}{\partial t} = 0$ , $\frac{\partial \vec{E}}{\partial t}=0$ and $\frac{\partial \vec{B}}{\partial t} = 0$,
the total electric plus magnetic torque would be:
\begin{align}
    \nabla \times \vec{F}_{E,M} & = \nabla \times \vec{F}_{E} + \nabla \times \vec{F}_{M}  \nonumber \\
     & = \nabla \rho_e \times \vec{E} + (\vec{B} \cdot \nabla) \vec{J} - (\vec{J} \cdot \nabla)\vec{B} ~.\label{cfEMstd}
\end{align}
In a conductive liquid, the current density $\vec{J}$ is related to both the electric field $\vec{E}$ and charge diffusion as given by the Einstein-Nernst equation,
\begin{equation}\label{EinsteinNerst}
    \vec{J} = \sigma \vec{E} - D \nabla \rho_e ~,
\end{equation}
where $\sigma$ and $D$ are conductivity and diffusion coefficients of the liquid, respectively. We can substitute  $\nabla \rho_e = \frac{1}{D}(\sigma \vec{E} - \vec{J})$ in Eq.~\ref{cfEMstd}; therefore,
\begin{equation}
    \nabla \times \vec{F}_{E,M} = \frac{1}{D} (\vec{E} \times \vec{J}) + (\vec{B} \cdot \nabla) \vec{J} - (\vec{J} \cdot \nabla) \vec{B} ~.\label{cfEMstd_1}
\end{equation}

This equation explicitly shows that there are only three terms in which the torque or rotational force appears. Furthermore, we assume that the electric/magnetic fields $\vec{E}$ and $\vec{B}$ in the right-hand side of the previous equation can be approximated by the external electric/magnetic fields $\vec{E}_{ext}$ and $\vec{B}_{ext}$, respectively; this implies that we will neglect the contributions from the electric and magnetic polarizations of the fluid in the first approximation. Therefore, Eqs.~\ref{2} and \ref{cfEMstd_1} are equal:
\begin{align}
   \frac{1}{D}(\vec{E} \times \vec{J}) + (\vec{B} \cdot \nabla) \vec{J} - (\vec{J} \cdot \nabla) \vec{B} & = (\vec{v} \cdot \nabla) \vec{\omega} - (\vec{\omega} \cdot \nabla) \vec{v} \nonumber \\ & - \nabla \times (\mu \nabla^{2} \vec{v}). \label{14}
\end{align}

\section{Experimental Apparatus}
\label{sec:expapparat}

Regarding the theoretical basis (see Eq.~\ref{14}),only three terms contribute to $\nabla \times \vec{F}_{EM}$. So one expects that only in this three cases, external electric or magnetic fields on current-carrying fluid produce a steady rotational flow. These three terms investigated in details experimentally as follows.

\subsection{Application of electric field in absence of magnetic field}

\begin{figure}
\includegraphics[width=9cm]{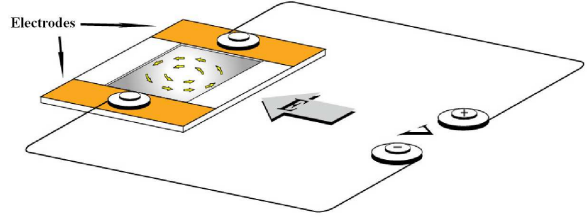}
 \caption{ Rotation of a suspended fluid film by electric fields. Magnetic field is absent.}
\label{fig1}
\end{figure}

In absence of magnetic field the second and third terms of equation \ref{cfEMstd_1} vanish. Therfore the torque is: $\nabla \times \vec{F} = \frac{1}{D}(\vec{E}_{ext} \times \vec{J})$. The effect of this term is reported in a study on `liquid film motor'\cite{amjadi2009,shirsavar2011}. Thin films always rotate in the direction of $\vec{E} \times \vec{J}$ and our previous experiments on different liquids implies that this simple rule is general. 
We use two electric power supplies to produce $\vec{J}$ and $\vec{E}_{ext}$. In experiments, in order to maxiaize the torque, the electric field and the electric current where choasen to be perpendicular ($\vec{E}_{ext} \perp \vec{J}$). The experimental set-up is shown in Fig~\ref{fig1}.

Contrary to the case for the liquid film, no rotation is observed in the bulk liquid; thus, the rotation is a surface effect. The reason for absence of rotation in bulk liquid is the short penetration depth of electric field in bulk liquid. In other word accumulation of charges on the interface between air and liquid causes the electric field to be screened.

\subsection{Application of magnetic field}

Here we investigate the case of \emph{bulk} liquid. When a magnetic field is present and the external electric field is zero ($\vec{B}_{ext} \neq 0$ and $\vec{E}_{ext} = 0$), one obtains for the bulk fluid,
\begin{equation}
\label{GammaBJ}
    \nabla \times \vec{F}_{EM}= (\vec{B}_{ext} \cdot \nabla) \vec{J} - (\vec{J} \cdot \nabla) \vec{B}_{ext} ~.
\end{equation}

Two different set-ups are designed to study the effect of the magnetic field. In both cases, the electrical conductivity of distilled water (used as the liquid) is increased to $13.37 ~ mS/cm$ by adding sodium chloride (NaCl): (\emph{i}) a uniform magnetic field $\vec{B}$ plus a non-uniform $\vec{J}$ are applied to the liquid; (\emph{ii}) a uniform current $\vec{J}$ plus a non-uniform $\vec{B}$ are applied to the liquid.

\begin{figure}
\begin{tabular}{cc}
{(a)} & \includegraphics[width=7.5 cm, angle=0]{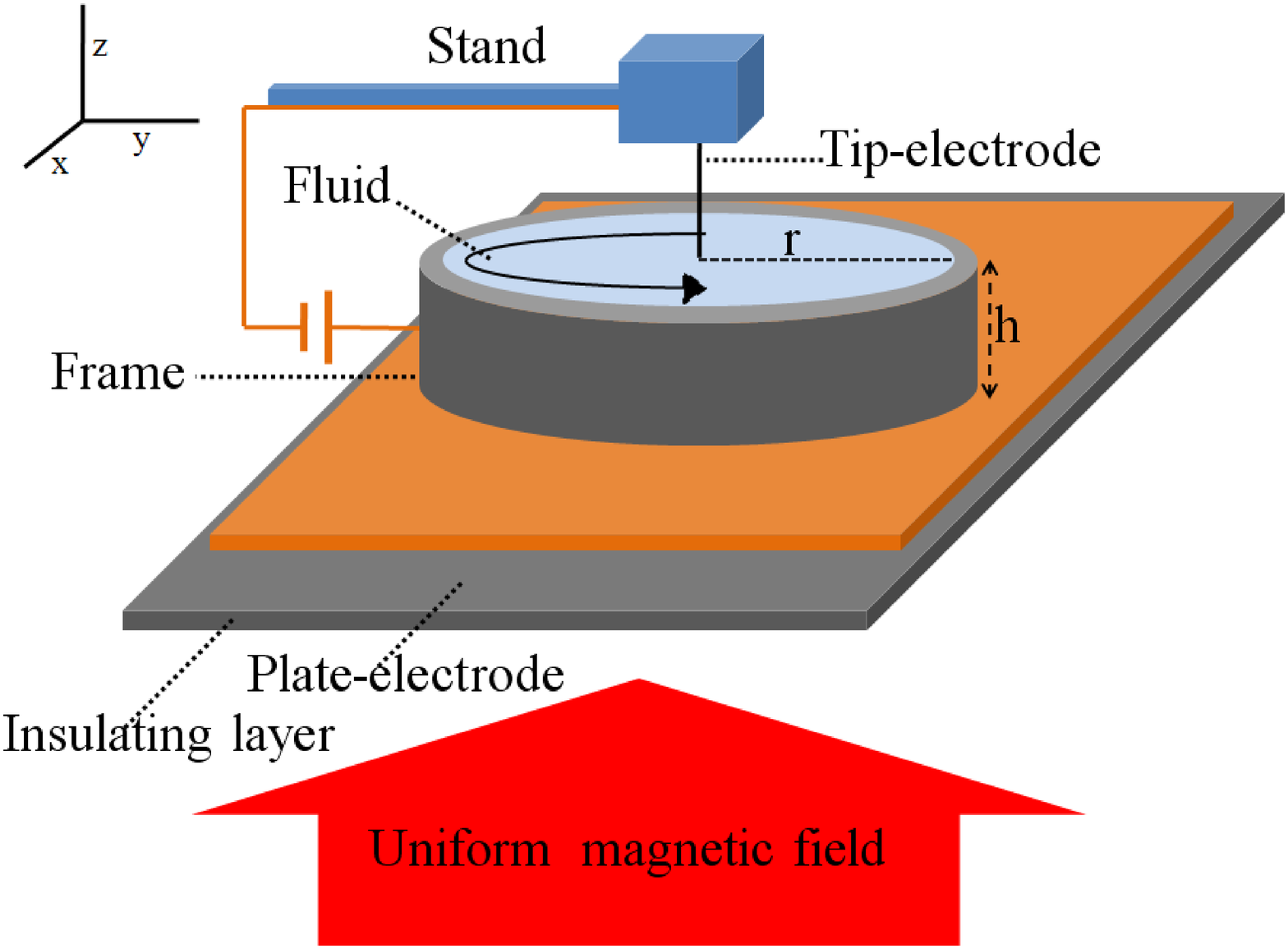}
\end{tabular}
\begin{tabular}{cccc}
{(b)} & \includegraphics[width=3.75 cm, angle=0]{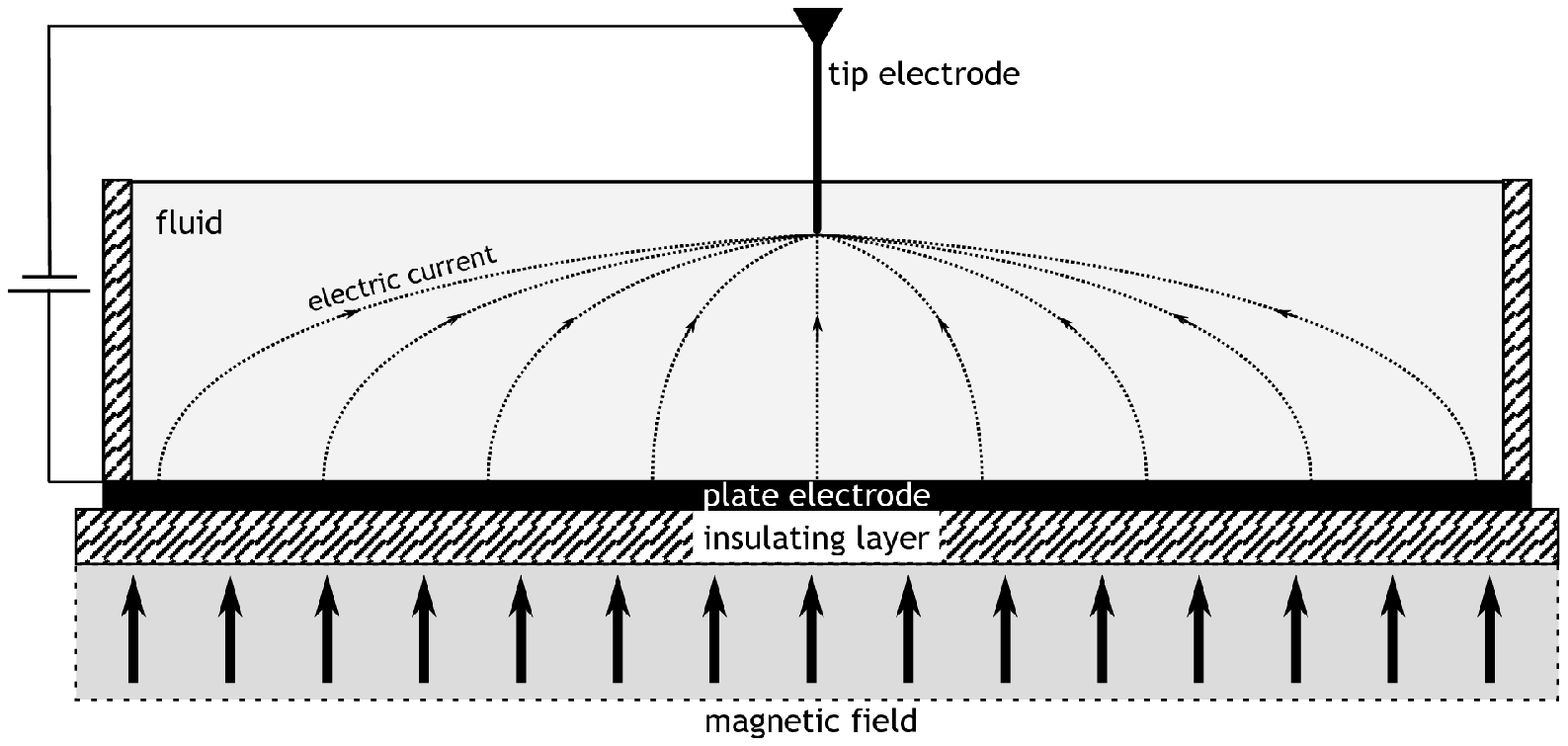} 
{(c)} & \includegraphics[width=2.75 cm, angle=0]{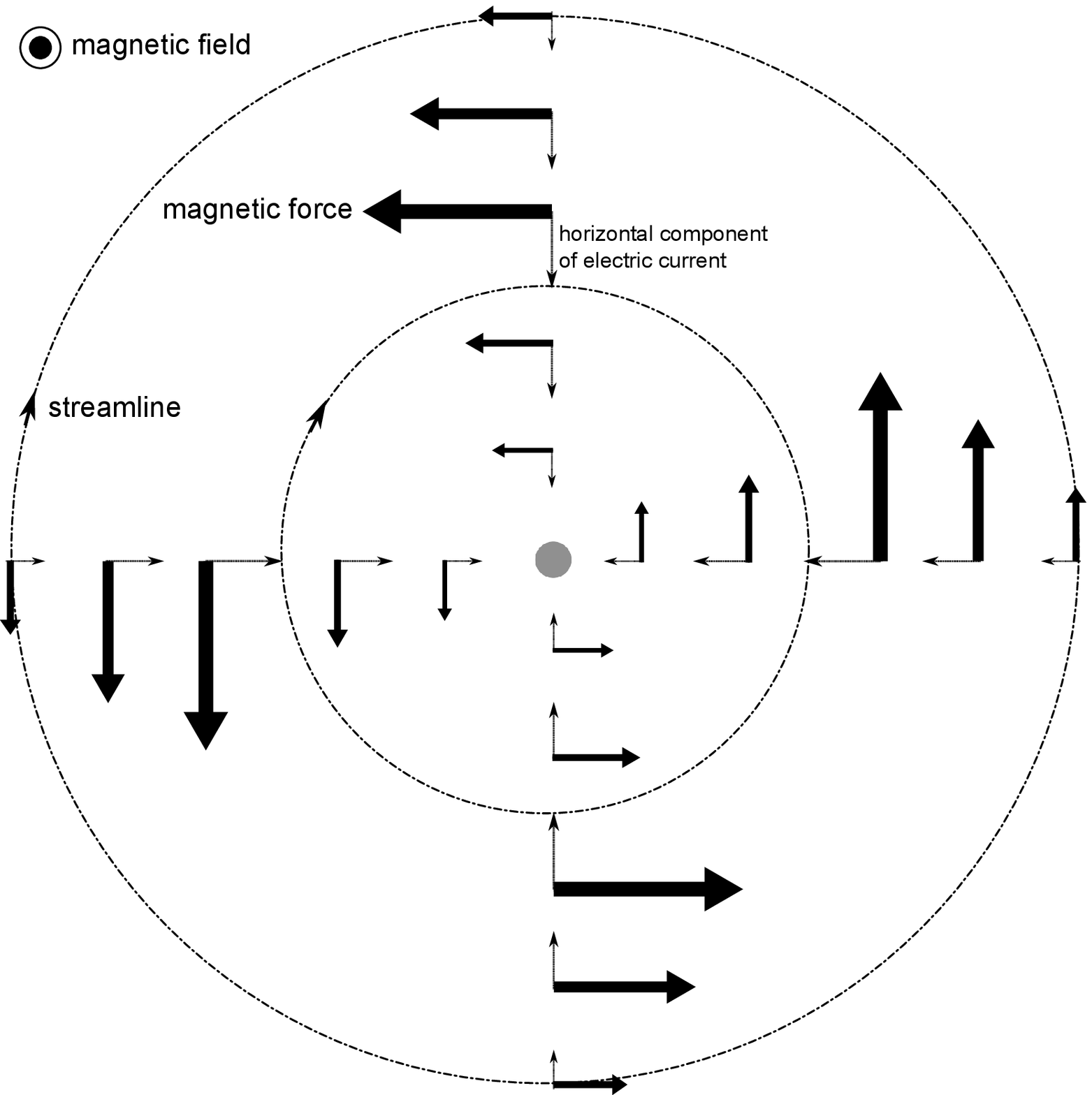} 
\end{tabular}
\begin{tabular}{cccc}
{(d)} & \includegraphics[width=3 cm, angle=0]{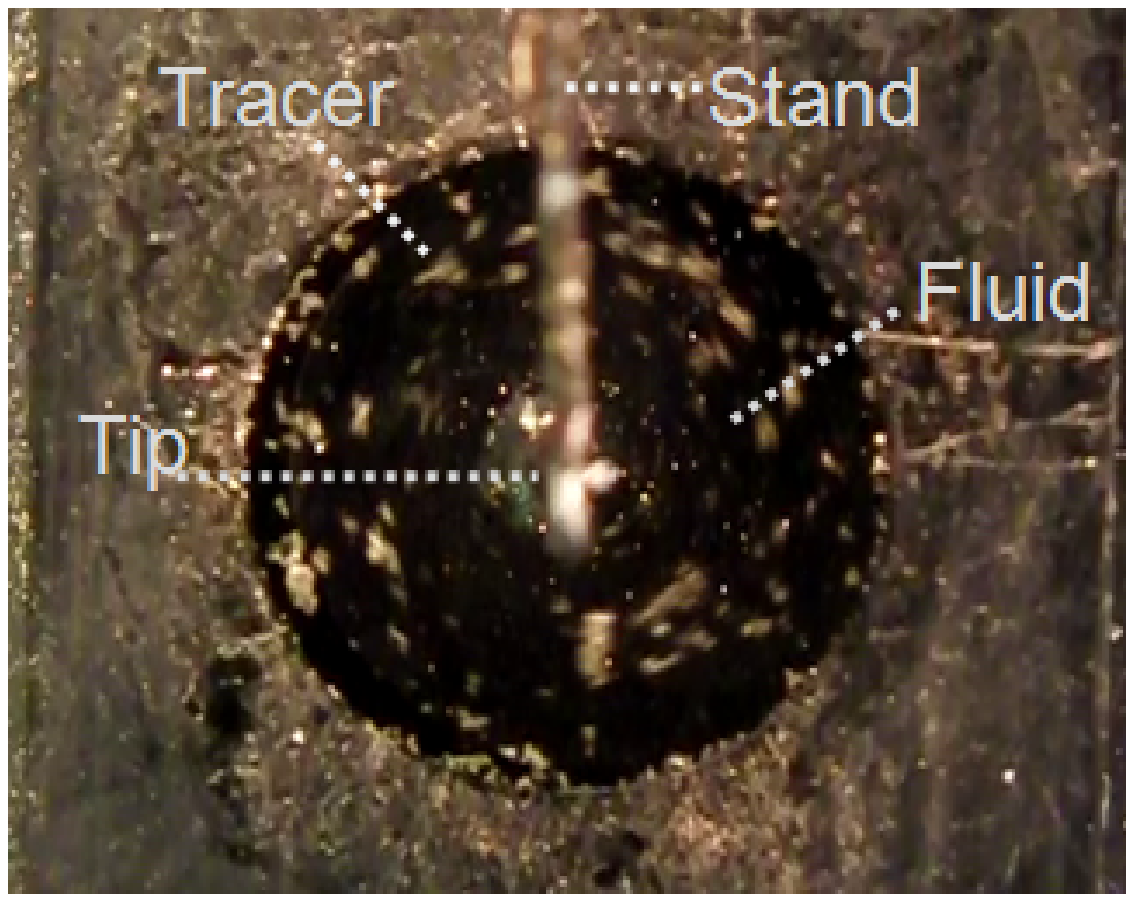} &
{(e)} & \includegraphics[width=3.5 cm, angle=0]{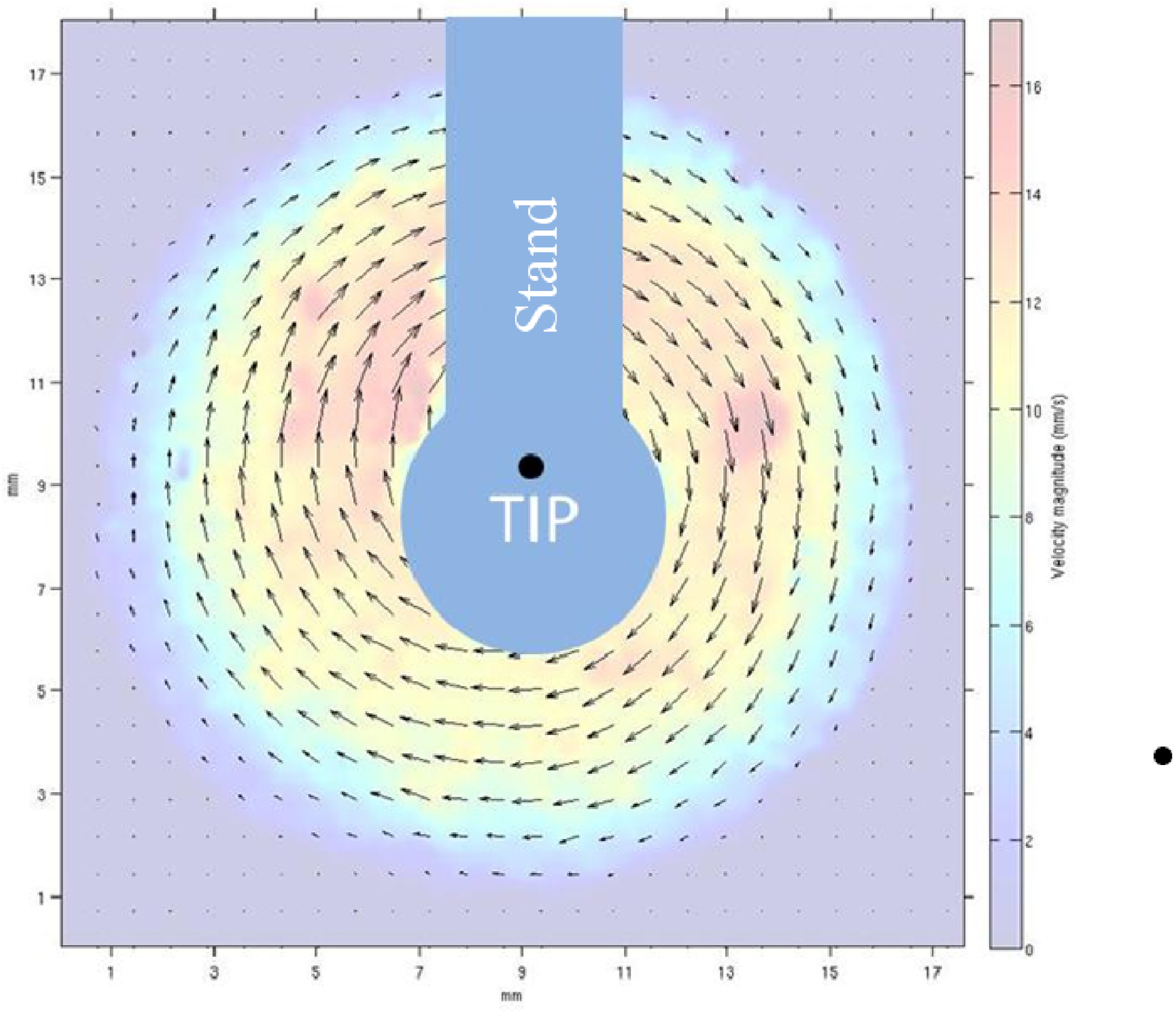} 
\end{tabular}
\caption{(a) Fluid bulk carrying an inhomogeneous current in presence of a uniform magnetic field. (b) Side view of the set-up: non-uniform electric current and the uniform magnetic field. (c) The amplitude of the magnetic force on the electric current is shown. (d) Photo of experimental set-up. (e) PIV pattern of rotation in the circular frame.} \label{fig2}
\end{figure}
\begin{figure}
\begin{tabular}{cc}
{(a)} & \includegraphics[width=7cm, angle=0]{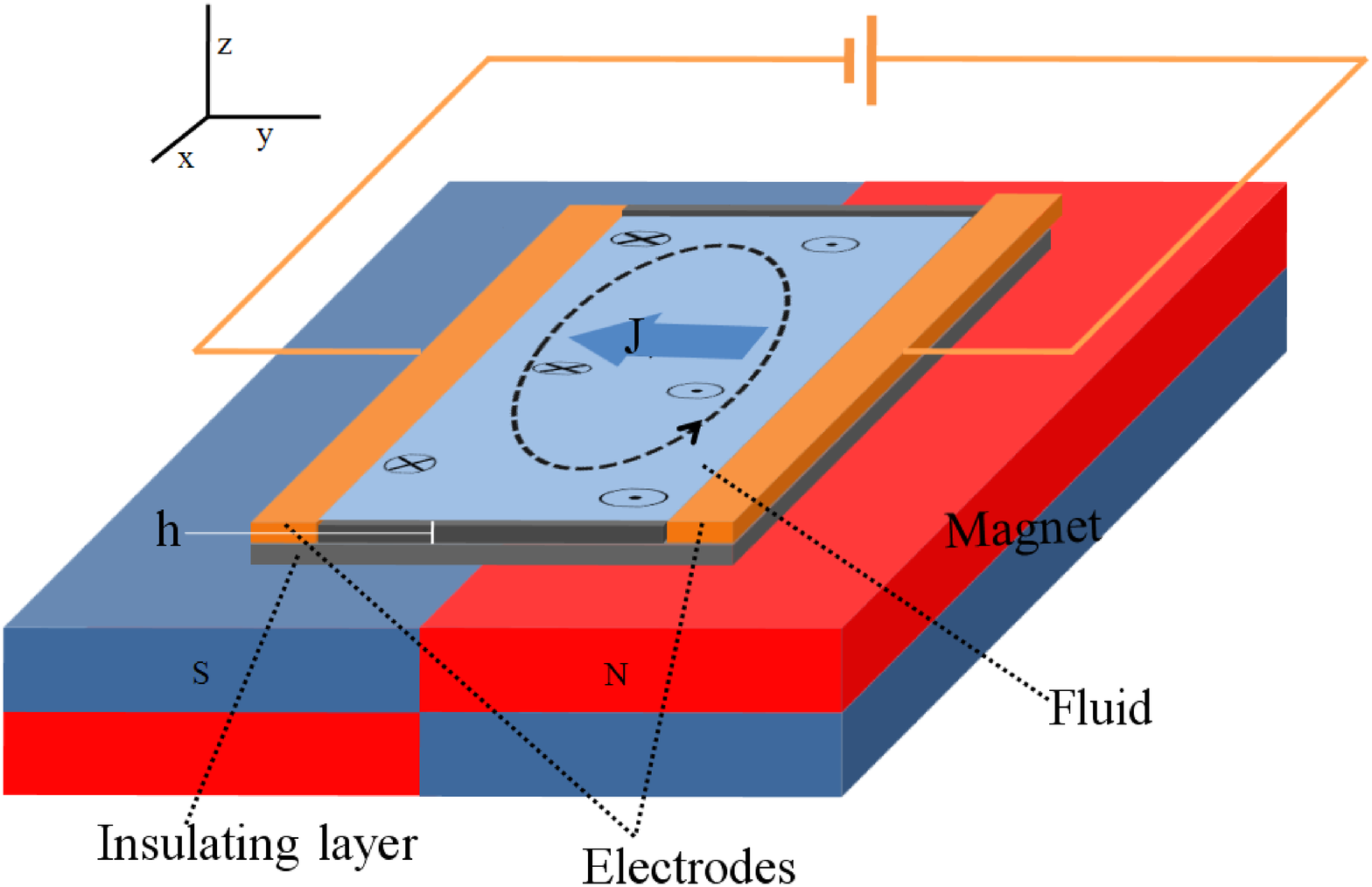}
\end{tabular}
\begin{tabular}{cccc}
{(b)} & \includegraphics[width=3.5 cm, angle=0]{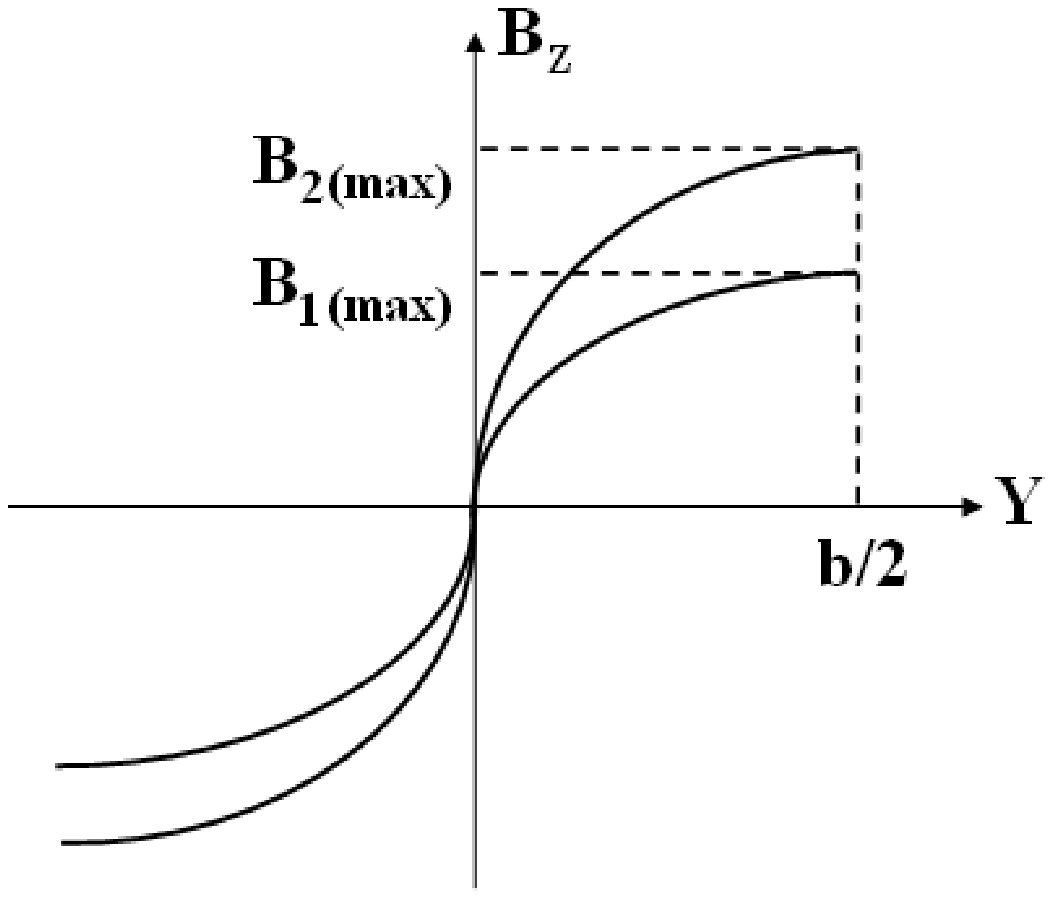} &
{(c)} & \includegraphics[width=2.5 cm, angle=0]{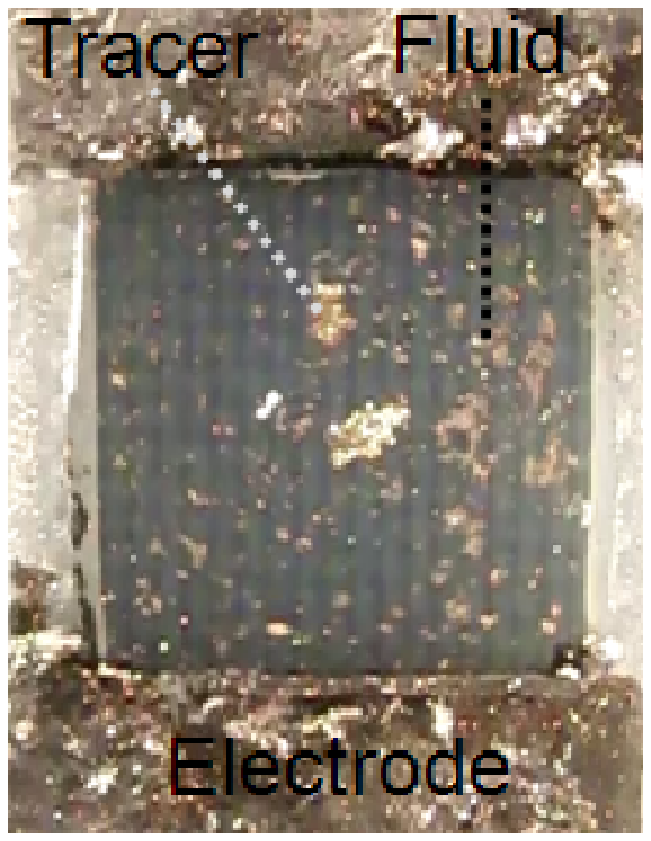} 
\end{tabular}
\begin{tabular}{cccc}
{(d)} & \includegraphics[width=3.5 cm, angle=0]{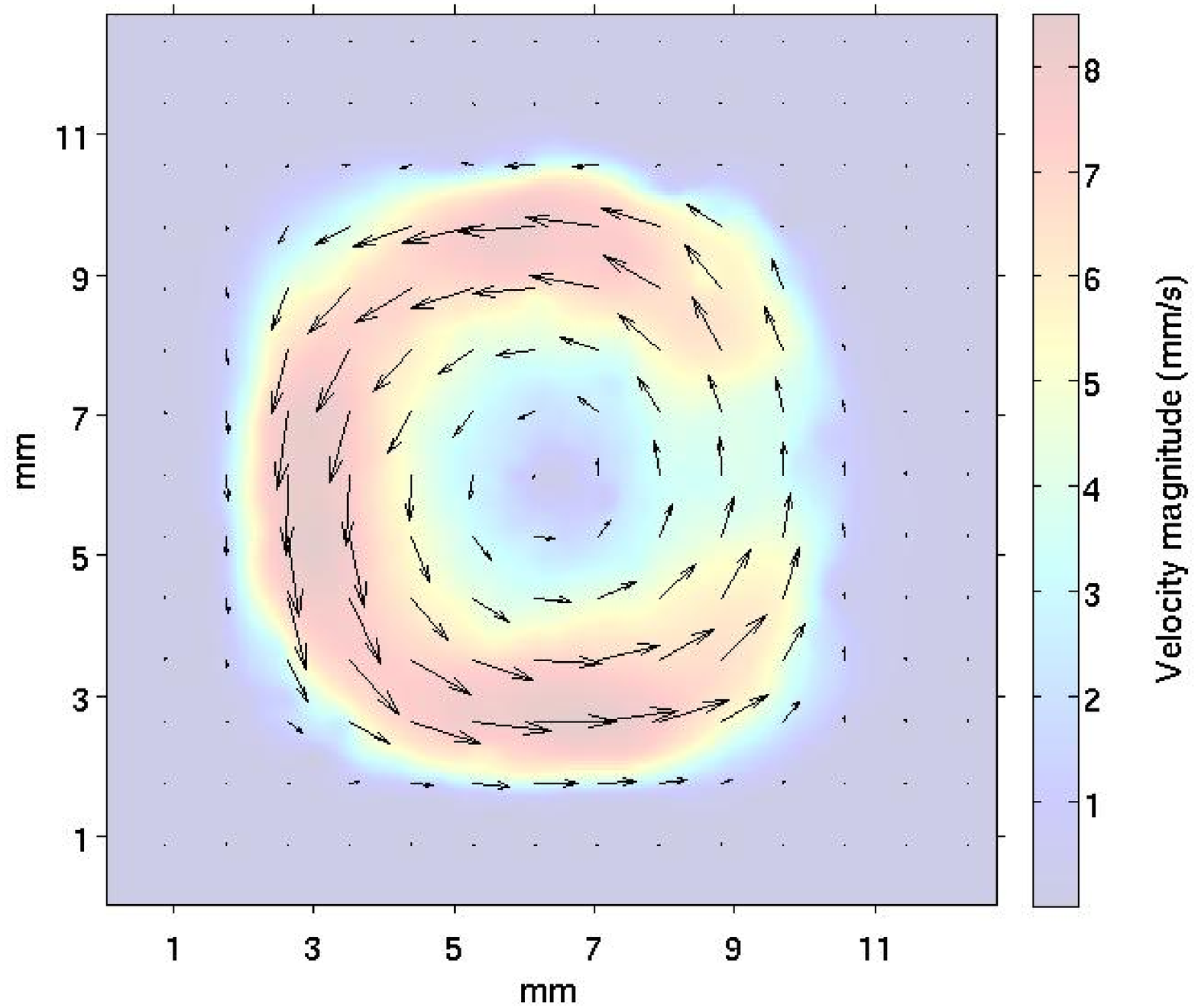} &
{(e)} & \includegraphics[width=2.5 cm, angle=0]{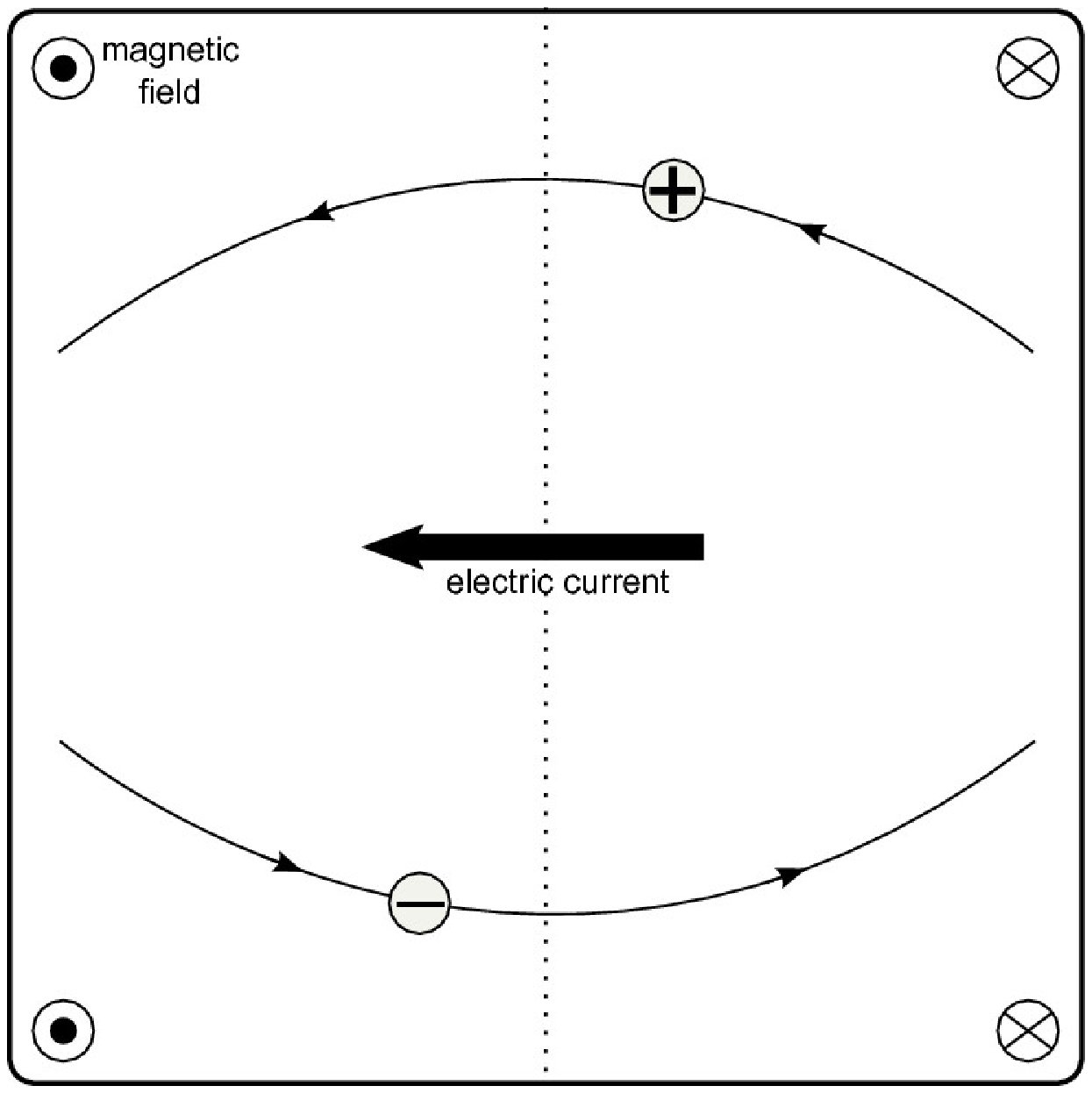} 
\end{tabular}
\caption{(a) Fluid bulk carrying a uniform current in presence of a non-uniform magnetic field. (b) The magnetic field is schematically shown along the $y$-direction. (c) A view of the cubic frame. (d) PIV pattern of rotation in rectangular frame. (e) The exerted force on ions in the non-uniform magnetic field} \label{fig3}
\end{figure}

\subsubsection{Uniform magnetic field and non-uniform electric current}
The set-up consists of a cylindrical cell ($\text{radius} = 7 ~ mm, \text{height} = 6 ~ mm$), with a plate-electrode at the bottom and a tip-electrode on top (Fig.~\ref{fig2}). In this way, we have a current gradient passing through the liquid.

We apply a uniform magnetic field in the vertical ($z$) direction and observe the flow of the fluid.
The velocity field of the observed rotation is obtained at different electric currents and two values of magnetic fields ($0.16 \pm 0.01 ~ T$ and $0.23 \pm 0.01 ~ T$). The data are polted in Fig~\ref{fig4}a,d. (For experiment movies see: \textit{http://portal.physics.sharif.edu/web/medphyslab/research})

\subsubsection{Non-uniform magnetic field and uniform electric current}

The set-up consists of a cubic cell ($xyz: ~ 1 \times 1 \times 0.1 ~ cm^{3}$) with two identical plate electrodes on its opposite sides. A uniform electric current passes through the fluid while a non-uniform magnetic field is applied to the cell~(Fig.~\ref{fig3}a). The magnetic field is projected onto the plane of fluid surface, schematically shown in Fig.~\ref{fig3}b. To produce the non-uniform magnetic field we use two cubic magnets as shown in Fig.~\ref{fig3}a.
The vertical ($z$) component of the magnetic field is upward in one side, while it is downward in the other side.

We measure the velocity pattern inside the cubic cell while a current of magnitude $0.3 ~ mA$ to $10 ~ mA$ is passing through it. In this experiment, the maximum magnitude of the magnetic field $B_0$ are $0.21 \pm 0.01 ~ T$ and $0.32 \pm 0.01 ~ T$.
To investigate the effect of electric current $J$ on the velocity field, we increase the applied electric current by increasing the voltage, and then perform velocimetry for each case. The data are plotted in Fig.\ref{fig4}(c,d) for the two values of magnetic fields.

\subsection{Velocimetry}
\label{sec:velocimetry}

After taking photos from tracer particles inside the liquid, Particle Image Velocimetry (PIV) was performed to plot the velocity vector field. The data are shown in Fig.~\ref{fig2}e and Fig.~\ref{fig3}d.

\section{Results and Discussion}
\label{sec:results}
Regarding equation~\ref{cfEMstd_1} there are only three terms in which $\nabla \times \vec{F} \neq 0$, and torque produces rotation flow.

\subsection{Application of electric field in the absence of magnetic field}
The first term $\frac{1}{D}(\vec{E} \times \vec{J})$ is surface effect.

Our experiments indicate that application of an electric current plus an electric field produces no observable steady flows in the fluid \emph{bulk}. The reason is the short penetration depth of electric field in conductive liquid. In contrast to the case when such a electric field and voltage are applied on a suspended fluid \emph{film}, rotation have been stuied on  water, polar liquids and suspended liquid crystal films \cite{amjadi2009,shirsavar2011,shirsavar2012}.
\subsection{Uniform magnetic field and non-uniform electric current}
The seccond term is an effect in bulk liquid.

By applying a uniform magnetic field on the cylindrical cell in the vertical ($z$) direction, and a current between the tip and plate electrodes, the fluid starts to rotate. Let us mention here that uniform magnetic field and uniform electric current could produce force but not torque .The average rotational velocity depends on the magnitude of the electric current and the magnetic field. The measured velocity pattern is shown in Fig.~\ref{fig2}e.
To justify this rotation, we consider the magnetic force exerted on the electric current ($\vec{J} \times \vec{B}_{ext}$). The current consists of a vertical $\vec{J}_{z}$ and a radial $\vec{J}_{r}$ component. The exerted force on $\vec{J}_{z}$ is zero, while $\vec{J}_{r} \times \vec{B}_{ext}$ produces a torque on the fluid (Fig.~\ref{fig2}) and a rotational flow appears on the liquid. 

\subsection{Non-uniform magnetic field and uniform electric current}
The third term is an effect in bulk liquid.

By applying a non-uniform magnetic field on the rectangular cell and establishing a uniform current, the liquid starts to rotate (Fig.~\ref{fig3}).
In this case, the magnetic field has two components, $\vec{B}_{\parallel}$ parallel to the current and $\vec{B}_{\perp}$ perpendicular to the current. 
The first component of force $\vec{J} \times \vec{B}_{\parallel}$ is zero while the second one $\vec{J} \times \vec{B}_{\perp}$ acts on the liquid, and causes the liquid to rotate. As shown in Fig.~\ref{fig3}e, regardless of the charge of the flowing particle, the exerted force is upward in the right-hand side and downward in the left-hand side. This can explain the sense of the rotation.

The plot of average velocity versus electric current (Fig.~\ref{fig4}(c,d)) confirms that the velocity increases when the applied voltage is increased.

\subsection{Angular velocity dependence on the electric current and magnetic field}

\begin{figure}[h!]
\begin{tabular}{cc}
{(a)} & \includegraphics[width=7 cm, angle=0]{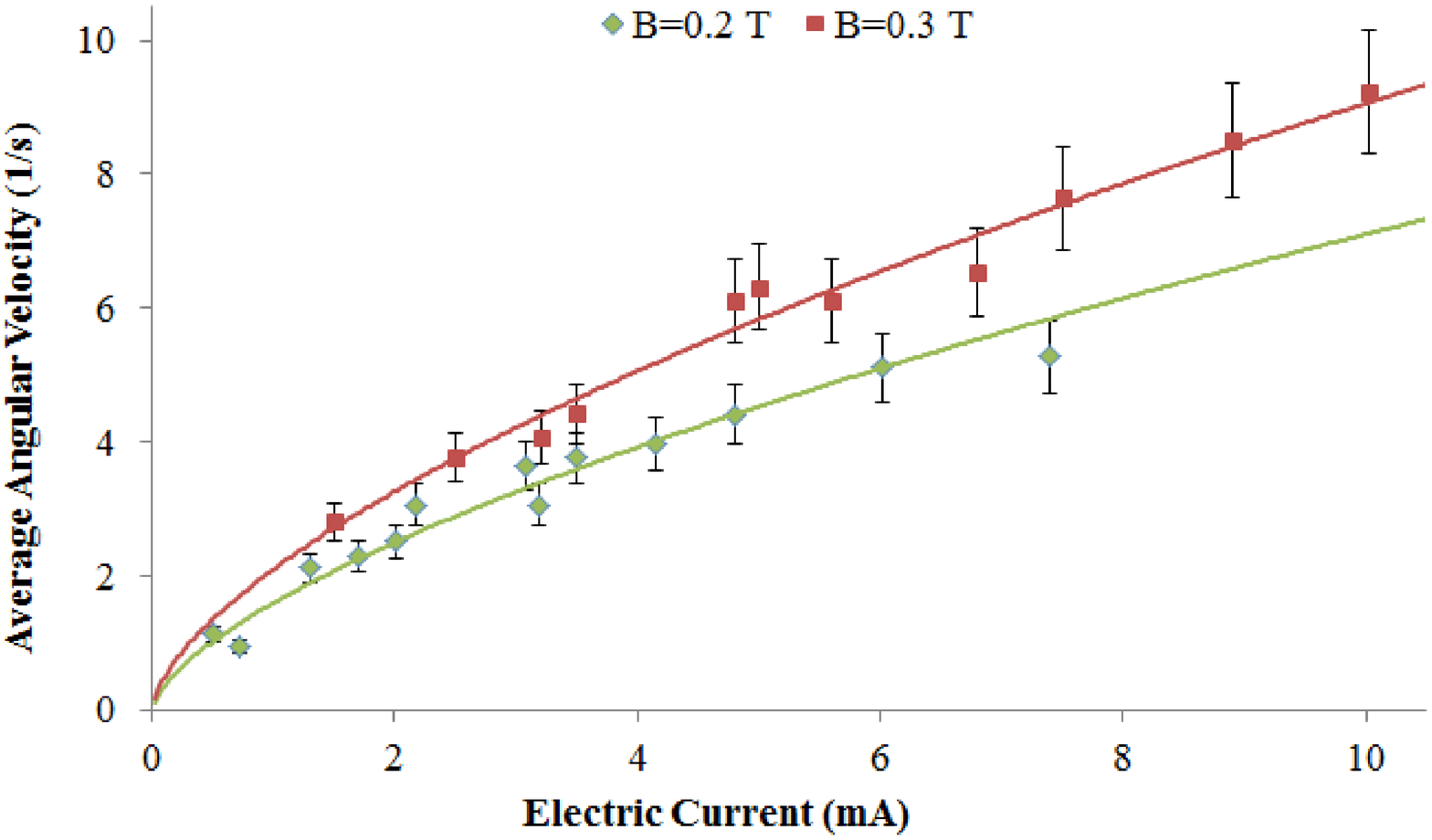} 
\end{tabular}
\begin{tabular}{cc}
{(b)} & \includegraphics[width=7 cm, angle=0]{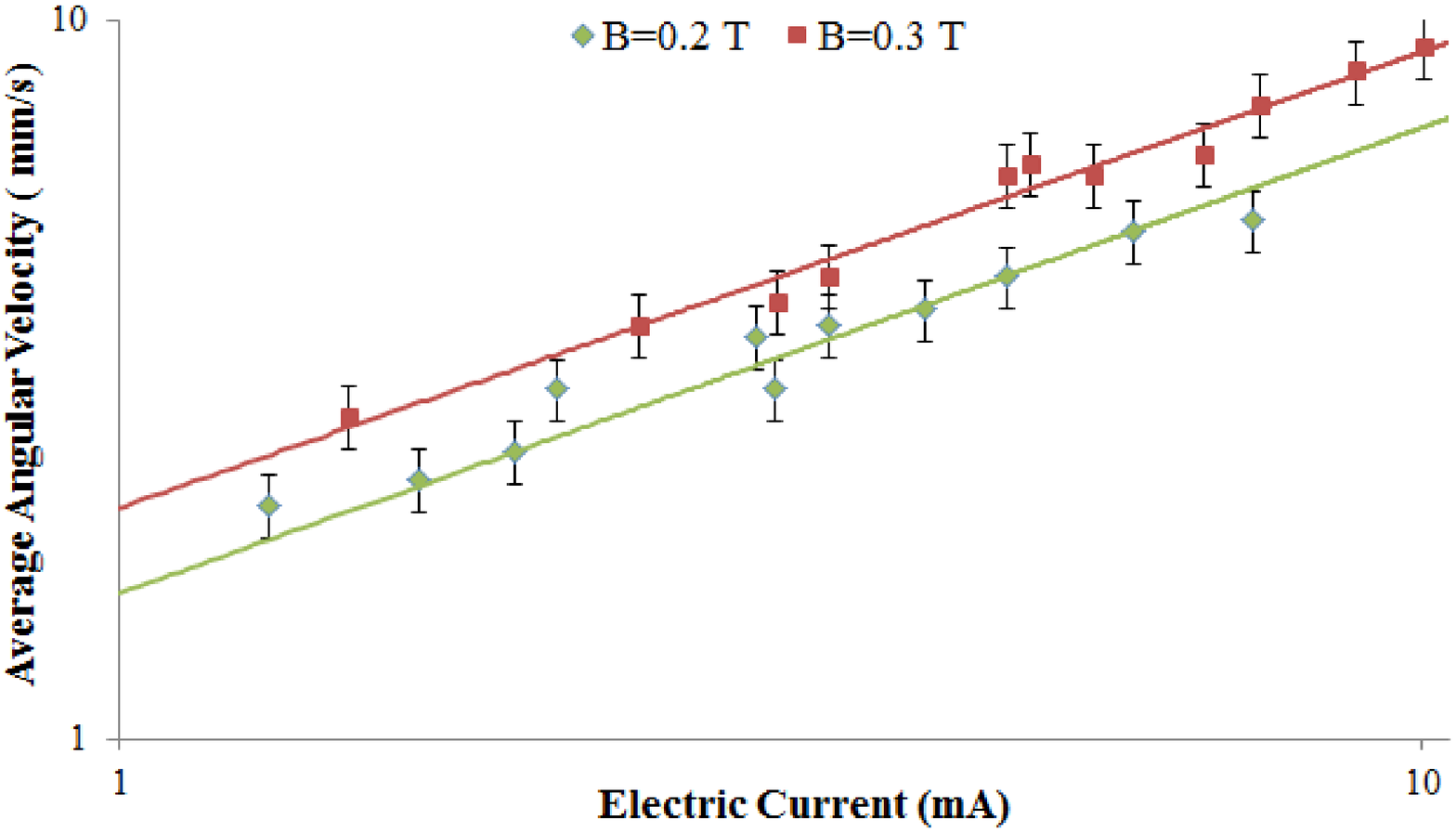}
\end{tabular}
\begin{tabular}{cc}
{(c)} & \includegraphics[width=7 cm, angle=0]{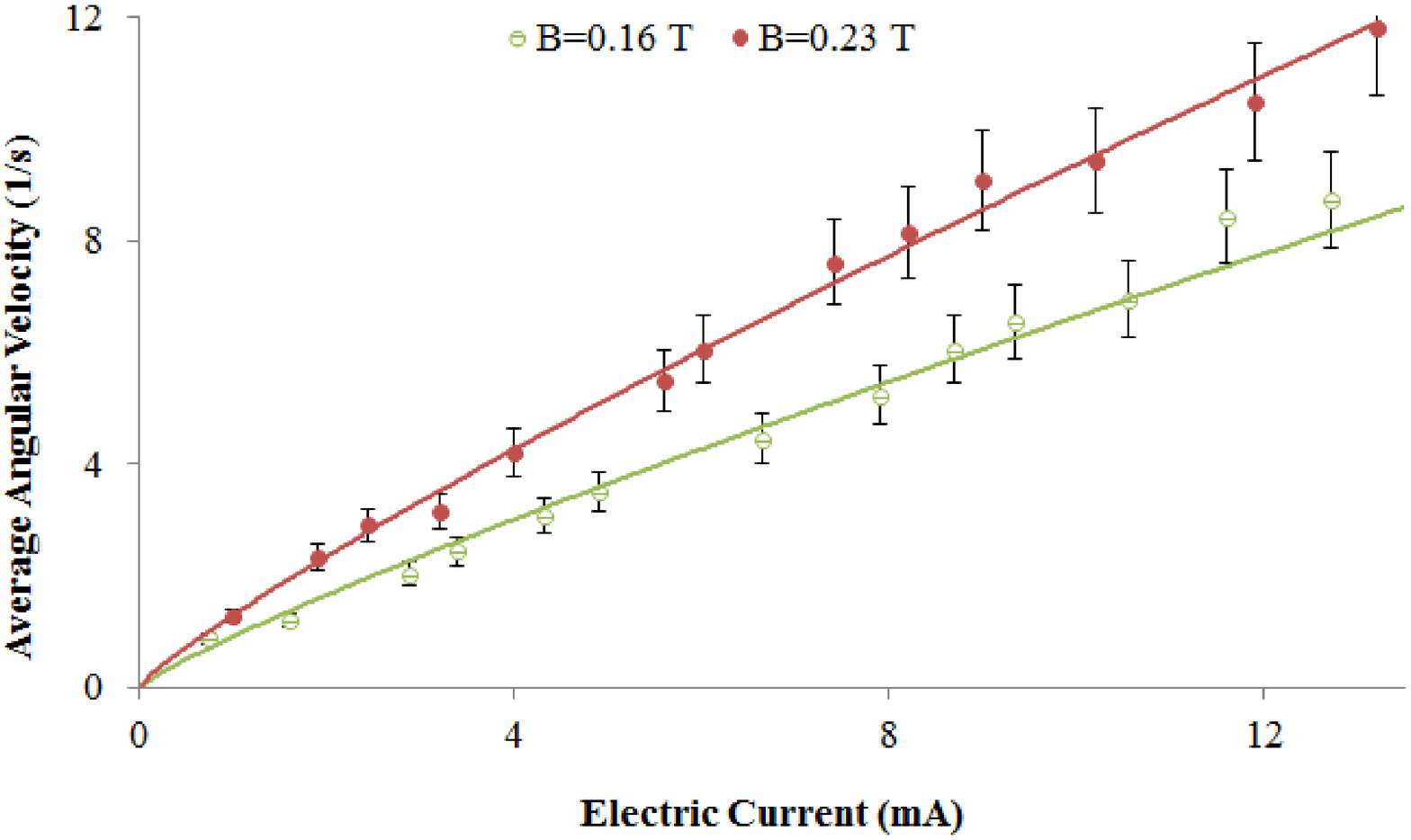}
\end{tabular}
\begin{tabular}{cc}
{(d)} & \includegraphics[width=7 cm, angle=0]{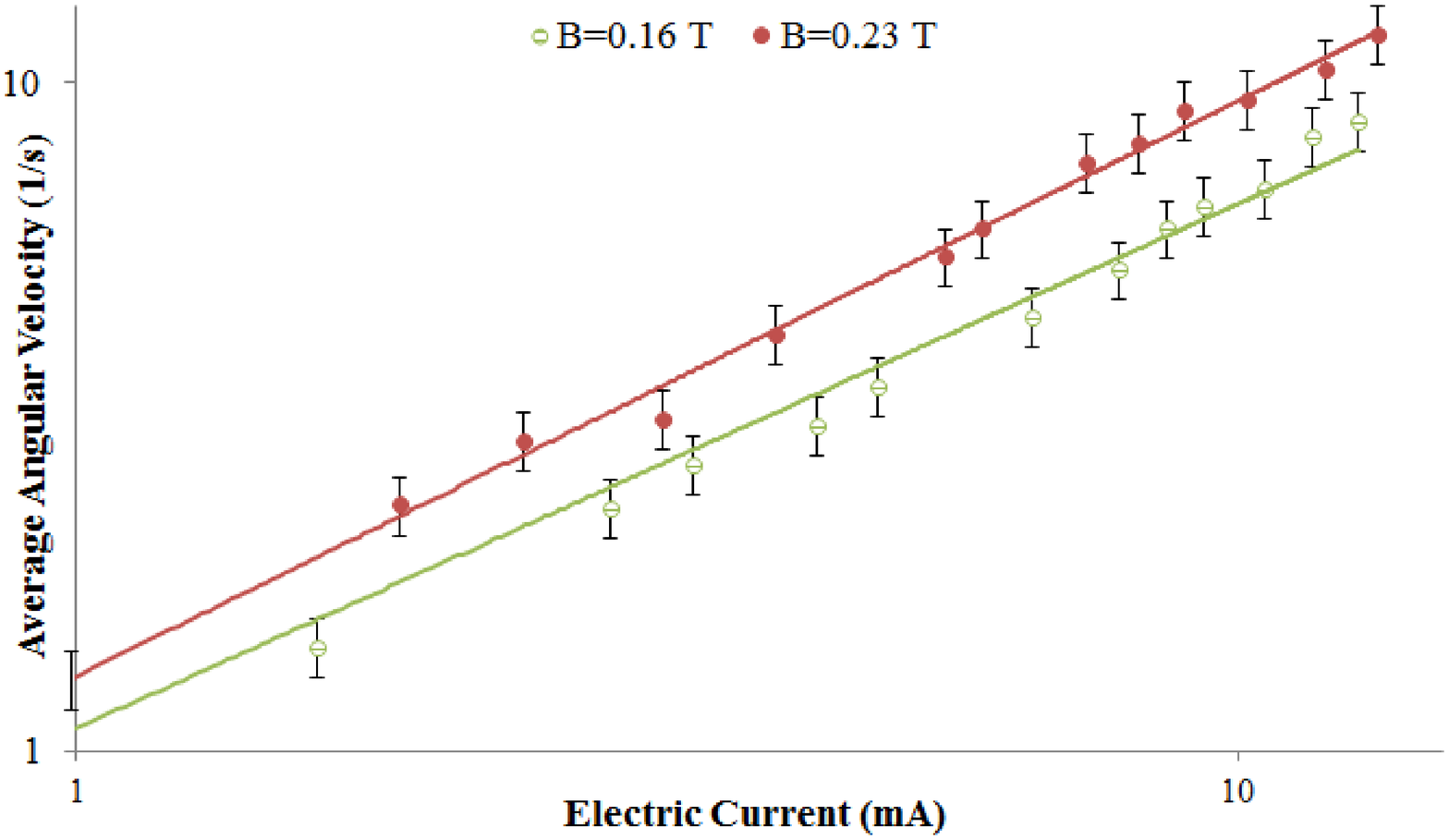} 
\end{tabular}
\caption{(a) Average angular velocity measurement versus electric current for the rectangular frame at a distance $2.5 ~ mm$ from the center of the rotation for two different magnetic values. (b) Log-log plot of average velocity versus electric current for the rectangular frame. (c) Average angular velocity versus electric current for the circular frame at a distance $3.5 ~ mm$ from the center of the rotation for two different magnetic values. (d) Log-log plot of average velocity versus electric current for the circular frame.} \label{fig4}
\end{figure}

We have used two different shapes of electrically insulating frames in our experiments: (\emph{i}) a rectangular frame, in order to apply a uniform electric current and non-uniform magnetic field to the liquid, and (\emph{ii}) a circular frame, to produce a uniform magnetic field and non-uniform electric current on liquid. Average angular velocity versus electric current and its log-log plot for two values of magnetic field are shown on Fig.~\ref{fig4}. These plots indicate two physical phenomena of non-linear systems. In a non-linear rotational flow, the dependence of angular velocity $\omega$ as a function of electric current $J$ should have the form $\omega = \beta J^{\alpha}$ where $\alpha$ is a positive value less than one indicating the non-linearity of the system. Taking the logarithm, one obtains $\log\omega = \log\beta + \alpha \log J$.

The plot is a line with constant slope $\alpha$ which shows that the average angular velocity should be a power function of $J$. From our experimental data for cylindrical and rectangular cells, these coefficients were determined to be $\alpha_{\circ} = 0.86$ for the cylindrical and $\alpha_{\square} = 0.63$ for the rectangular cell. This implies that in the cylindrical cell less energy is required to produce rotational flow with a given $\omega$, than in a rectangular cell with similar dimensions. Experimental data for two different magnetic fields $B$ (Fig.~\ref{fig4}.c and Fig.~\ref{fig4}.d) shows that for higher magnetic fields the plot starts from higher values ($\log\beta$). This simply implies that higher magnetic fields produce stronger forces on the fluid and consequently, it rotates with higher velocities.

\section{Conclusion}

We have studied the rotational behaviour of an electrically conductive liquid under the application of external electric/magnetic fields with different configurations. A fundamental theoretical basis supports the experimental set-ups in which we have observed the effect of rotational forces or toque (for which $\nabla \times \vec{F} \neq 0$). 
The theoretical basis suggests three general cases:
\begin{enumerate}
 \item A \emph{constant} electric field perpendicular to an electric current on the liquid film, while magnetic field is zero. This is a surface effect.
 \item A \emph{constant} magnetic field and a \emph{non-uniform} electric current are applied in liquid bulk. The fluid rotates in the direction of $(\vec{B} \cdot \nabla) \vec{J} $ and the average rotational velocity increases by increasing the current or the magnetic field.
 \item A \emph{non-uniform} magnetic field and a \emph{constant} electric current are applied in liquid bulk. The fluid rotates in the direction of $-( \vec{J} \cdot \nabla) \vec{B} $ and the average rotational velocity increases with increasing current or magnetic field.
\end{enumerate}

In the presence of a magnetic field from a log-log plot of average angular velocity of a conductive laminar flow through a porous medium versus the applied current, one can compute $\alpha_{0}$ and determine the porosity of the medium.

As the experimental methods and results are applicable to fairly diverse situations, especially to the bulk fluid, they could be used to develop novel methods or designs for manipulation of fluid flow in many important areas 
including biology and small-scale industry, \emph{e.g.} 'lab-on-chip' devices.

\section*{Acknowledgement}
This work is performed in Medical Physics and Laser Lab supported by Sharif University Applied Physics Research Center. We would like to acknowledge the fruitful discussions with Prof. M. Zahn and Prof. M.R. Ejtehadi and also, the technical help from Dr M. R. Mozaffari and M. K. Zand.

\bibliography{bib}

\begin{thebibliography}{21}%
\makeatletter
\providecommand \@ifxundefined [1]{%
 \@ifx{#1\undefined}
}%
\providecommand \@ifnum [1]{%
 \ifnum #1\expandafter \@firstoftwo
 \else \expandafter \@secondoftwo
 \fi
}%
\providecommand \@ifx [1]{%
 \ifx #1\expandafter \@firstoftwo
 \else \expandafter \@secondoftwo
 \fi
}%
\providecommand \natexlab [1]{#1}%
\providecommand \enquote  [1]{``#1''}%
\providecommand \bibnamefont  [1]{#1}%
\providecommand \bibfnamefont [1]{#1}%
\providecommand \citenamefont [1]{#1}%
\providecommand \href@noop [0]{\@secondoftwo}%
\providecommand \href [0]{\begingroup \@sanitize@url \@href}%
\providecommand \@href[1]{\@@startlink{#1}\@@href}%
\providecommand \@@href[1]{\endgroup#1\@@endlink}%
\providecommand \@sanitize@url [0]{\catcode `\\12\catcode `\$12\catcode
  `\&12\catcode `\#12\catcode `\^12\catcode `\_12\catcode `\%12\relax}%
\providecommand \@@startlink[1]{}%
\providecommand \@@endlink[0]{}%
\providecommand \url  [0]{\begingroup\@sanitize@url \@url }%
\providecommand \@url [1]{\endgroup\@href {#1}{\urlprefix }}%
\providecommand \urlprefix  [0]{URL }%
\providecommand \Eprint [0]{\href }%
\providecommand \doibase [0]{http://dx.doi.org/}%
\providecommand \selectlanguage [0]{\@gobble}%
\providecommand \bibinfo  [0]{\@secondoftwo}%
\providecommand \bibfield  [0]{\@secondoftwo}%
\providecommand \translation [1]{[#1]}%
\providecommand \BibitemOpen [0]{}%
\providecommand \bibitemStop [0]{}%
\providecommand \bibitemNoStop [0]{.\EOS\space}%
\providecommand \EOS [0]{\spacefactor3000\relax}%
\providecommand \BibitemShut  [1]{\csname bibitem#1\endcsname}%
\let\auto@bib@innerbib\@empty
\bibitem [{\citenamefont {Melcher}(1961)}]{melcher1961}%
  \BibitemOpen
  \bibfield  {author} {\bibinfo {author} {\bibfnamefont {J.~R.}\ \bibnamefont
  {Melcher}},\ }\bibfield  {title} {\enquote {\bibinfo {title}
  {Electrohydrodynamic and magnetohydrodynamic surface waves and
  instabilities},}\ }\href@noop {} {\bibfield  {journal} {\bibinfo  {journal}
  {Physics of Fluids (US)}\ }\textbf {\bibinfo {volume} {4}} (\bibinfo {year}
  {1961})}\BibitemShut {NoStop}%
\bibitem [{\citenamefont {Melcher}\ and\ \citenamefont
  {Smith}(1969)}]{melcher1969}%
  \BibitemOpen
  \bibfield  {author} {\bibinfo {author} {\bibfnamefont {J.~R.}\ \bibnamefont
  {Melcher}}\ and\ \bibinfo {author} {\bibfnamefont {C.~V.}\ \bibnamefont
  {Smith}},\ }\bibfield  {title} {\enquote {\bibinfo {title}
  {Electrohydrodynamic charge relaxation and interfacial perpendicular-field
  instability},}\ }\href@noop {} {\bibfield  {journal} {\bibinfo  {journal}
  {Physics of Fluids}\ }\textbf {\bibinfo {volume} {12}},\ \bibinfo {pages}
  {778--790} (\bibinfo {year} {1969})}\BibitemShut {NoStop}%
\bibitem [{\citenamefont {Washabauch}, \citenamefont {Zahn},\ and\
  \citenamefont {Melcher}(1989)}]{washabauch1989}%
  \BibitemOpen
  \bibfield  {author} {\bibinfo {author} {\bibfnamefont {A.}~\bibnamefont
  {Washabauch}}, \bibinfo {author} {\bibfnamefont {M.}~\bibnamefont {Zahn}}, \
  and\ \bibinfo {author} {\bibfnamefont {J.}~\bibnamefont {Melcher}},\
  }\bibfield  {title} {\enquote {\bibinfo {title} {Electrohydrodynamic
  traveling-wave pumping of homogeneous semi-insulating liquids},}\ }\href@noop
  {} {\bibfield  {journal} {\bibinfo  {journal} {Electrical Insulation, IEEE
  Transactions on}\ }\textbf {\bibinfo {volume} {24}},\ \bibinfo {pages}
  {807--834} (\bibinfo {year} {1989})}\BibitemShut {NoStop}%
\bibitem [{\citenamefont {Taylor}\ and\ \citenamefont
  {Taylor}(1966)}]{taylor1966}%
  \BibitemOpen
  \bibfield  {author} {\bibinfo {author} {\bibfnamefont {G.}~\bibnamefont
  {Taylor}}\ and\ \bibinfo {author} {\bibfnamefont {G.}~\bibnamefont
  {Taylor}},\ }\bibfield  {title} {\enquote {\bibinfo {title} {Studies in
  electrohydrodynamics. i. the circulation produced in a drop by electrical
  field},}\ }\href@noop {} {\bibfield  {journal} {\bibinfo  {journal}
  {Proceedings of the Royal Society of London. Series A. Mathematical and
  Physical Sciences}\ }\textbf {\bibinfo {volume} {291}},\ \bibinfo {pages}
  {159--166} (\bibinfo {year} {1966})}\BibitemShut {NoStop}%
\bibitem [{\citenamefont {Saville}(1997)}]{saville1997}%
  \BibitemOpen
  \bibfield  {author} {\bibinfo {author} {\bibfnamefont {D.}~\bibnamefont
  {Saville}},\ }\bibfield  {title} {\enquote {\bibinfo {title}
  {Electrohydrodynamics: the taylor-melcher leaky dielectric model},}\
  }\href@noop {} {\bibfield  {journal} {\bibinfo  {journal} {Annual review of
  fluid mechanics}\ }\textbf {\bibinfo {volume} {29}},\ \bibinfo {pages}
  {27--64} (\bibinfo {year} {1997})}\BibitemShut {NoStop}%
\bibitem [{\citenamefont {Ramos}\ \emph
  {et~al.}(1999{\natexlab{a}})\citenamefont {Ramos}, \citenamefont {Morgan},
  \citenamefont {Green},\ and\ \citenamefont {Castellanos}}]{ramos1999ac}%
  \BibitemOpen
  \bibfield  {author} {\bibinfo {author} {\bibfnamefont {A.}~\bibnamefont
  {Ramos}}, \bibinfo {author} {\bibfnamefont {H.}~\bibnamefont {Morgan}},
  \bibinfo {author} {\bibfnamefont {N.}~\bibnamefont {Green}}, \ and\ \bibinfo
  {author} {\bibfnamefont {A.}~\bibnamefont {Castellanos}},\ }\bibfield
  {title} {\enquote {\bibinfo {title} {Ac electrokinetics: a review of forces
  in microelectrode structures},}\ }\href@noop {} {\bibfield  {journal}
  {\bibinfo  {journal} {Journal of Physics D: Applied Physics}\ }\textbf
  {\bibinfo {volume} {31}},\ \bibinfo {pages} {2338} (\bibinfo {year}
  {1999}{\natexlab{a}})}\BibitemShut {NoStop}%
\bibitem [{\citenamefont {Ramos}\ \emph
  {et~al.}(1999{\natexlab{b}})\citenamefont {Ramos}, \citenamefont {Morgan},
  \citenamefont {Green},\ and\ \citenamefont {Castellanos}}]{ramos1999ab}%
  \BibitemOpen
  \bibfield  {author} {\bibinfo {author} {\bibfnamefont {A.}~\bibnamefont
  {Ramos}}, \bibinfo {author} {\bibfnamefont {H.}~\bibnamefont {Morgan}},
  \bibinfo {author} {\bibfnamefont {N.~G.}\ \bibnamefont {Green}}, \ and\
  \bibinfo {author} {\bibfnamefont {A.}~\bibnamefont {Castellanos}},\
  }\bibfield  {title} {\enquote {\bibinfo {title} {Ac electric-field-induced
  fluid flow in microelectrodes},}\ }\href@noop {} {\bibfield  {journal}
  {\bibinfo  {journal} {Journal of colloid and interface science}\ }\textbf
  {\bibinfo {volume} {217}},\ \bibinfo {pages} {420--422} (\bibinfo {year}
  {1999}{\natexlab{b}})}\BibitemShut {NoStop}%
\bibitem [{\citenamefont {Faetti}, \citenamefont {Fronzoni},\ and\
  \citenamefont {Rolla}(1983)}]{faetti1983}%
  \BibitemOpen
  \bibfield  {author} {\bibinfo {author} {\bibfnamefont {S.}~\bibnamefont
  {Faetti}}, \bibinfo {author} {\bibfnamefont {L.}~\bibnamefont {Fronzoni}}, \
  and\ \bibinfo {author} {\bibfnamefont {P.}~\bibnamefont {Rolla}},\ }\bibfield
   {title} {\enquote {\bibinfo {title} {Static and dynamic behavior of the
  vortex--electrohydrodynamic instability in freely suspended layers of nematic
  liquid crystals},}\ }\href@noop {} {\bibfield  {journal} {\bibinfo  {journal}
  {The Journal of chemical physics}\ }\textbf {\bibinfo {volume} {79}},\
  \bibinfo {pages} {5054} (\bibinfo {year} {1983})}\BibitemShut {NoStop}%
\bibitem [{\citenamefont {Faetti}, \citenamefont {Fronzoni},\ and\
  \citenamefont {Rolla}(1979)}]{faetti1979}%
  \BibitemOpen
  \bibfield  {author} {\bibinfo {author} {\bibfnamefont {S.}~\bibnamefont
  {Faetti}}, \bibinfo {author} {\bibfnamefont {L.}~\bibnamefont {Fronzoni}}, \
  and\ \bibinfo {author} {\bibfnamefont {P.}~\bibnamefont {Rolla}},\ }\bibfield
   {title} {\enquote {\bibinfo {title} {Electrohydrodynamic flow in nematic
  thin films with two free surfaces},}\ }\href@noop {} {\bibfield  {journal}
  {\bibinfo  {journal} {Le Journal de Physique Colloques}\ }\textbf {\bibinfo
  {volume} {40}},\ \bibinfo {pages} {3--3} (\bibinfo {year}
  {1979})}\BibitemShut {NoStop}%
\bibitem [{\citenamefont {Morris}, \citenamefont {de~Bruyn},\ and\
  \citenamefont {May}(1990)}]{morris1990}%
  \BibitemOpen
  \bibfield  {author} {\bibinfo {author} {\bibfnamefont {S.~W.}\ \bibnamefont
  {Morris}}, \bibinfo {author} {\bibfnamefont {J.~R.}\ \bibnamefont
  {de~Bruyn}}, \ and\ \bibinfo {author} {\bibfnamefont {A.}~\bibnamefont
  {May}},\ }\bibfield  {title} {\enquote {\bibinfo {title} {Electroconvection
  and pattern formation in a suspended smectic film},}\ }\href@noop {}
  {\bibfield  {journal} {\bibinfo  {journal} {Physical review letters}\
  }\textbf {\bibinfo {volume} {65}},\ \bibinfo {pages} {2378--2381} (\bibinfo
  {year} {1990})}\BibitemShut {NoStop}%
\bibitem [{\citenamefont {Daya}, \citenamefont {Morris},\ and\ \citenamefont
  {de~Bruyn}(1997)}]{daya1997}%
  \BibitemOpen
  \bibfield  {author} {\bibinfo {author} {\bibfnamefont {Z.~A.}\ \bibnamefont
  {Daya}}, \bibinfo {author} {\bibfnamefont {S.~W.}\ \bibnamefont {Morris}}, \
  and\ \bibinfo {author} {\bibfnamefont {J.~R.}\ \bibnamefont {de~Bruyn}},\
  }\bibfield  {title} {\enquote {\bibinfo {title} {Electroconvection in a
  suspended fluid film: a linear stability analysis},}\ }\href@noop {}
  {\bibfield  {journal} {\bibinfo  {journal} {Physical Review E}\ }\textbf
  {\bibinfo {volume} {55}},\ \bibinfo {pages} {2682} (\bibinfo {year}
  {1997})}\BibitemShut {NoStop}%
\bibitem [{\citenamefont {Amjadi}\ \emph {et~al.}(2009)\citenamefont {Amjadi},
  \citenamefont {Shirsavar}, \citenamefont {Radja},\ and\ \citenamefont
  {Ejtehadi}}]{amjadi2009}%
  \BibitemOpen
  \bibfield  {author} {\bibinfo {author} {\bibfnamefont {A.}~\bibnamefont
  {Amjadi}}, \bibinfo {author} {\bibfnamefont {R.}~\bibnamefont {Shirsavar}},
  \bibinfo {author} {\bibfnamefont {N.~H.}\ \bibnamefont {Radja}}, \ and\
  \bibinfo {author} {\bibfnamefont {M.}~\bibnamefont {Ejtehadi}},\ }\bibfield
  {title} {\enquote {\bibinfo {title} {A liquid film motor},}\ }\href@noop {}
  {\bibfield  {journal} {\bibinfo  {journal} {Microfluidics and nanofluidics}\
  }\textbf {\bibinfo {volume} {6}},\ \bibinfo {pages} {711--715} (\bibinfo
  {year} {2009})}\BibitemShut {NoStop}%
\bibitem [{\citenamefont {Shirsavar}\ \emph {et~al.}(2011)\citenamefont
  {Shirsavar}, \citenamefont {Amjadi}, \citenamefont {Tonddast-Navaei},\ and\
  \citenamefont {Ejtehadi}}]{shirsavar2011}%
  \BibitemOpen
  \bibfield  {author} {\bibinfo {author} {\bibfnamefont {R.}~\bibnamefont
  {Shirsavar}}, \bibinfo {author} {\bibfnamefont {A.}~\bibnamefont {Amjadi}},
  \bibinfo {author} {\bibfnamefont {A.}~\bibnamefont {Tonddast-Navaei}}, \ and\
  \bibinfo {author} {\bibfnamefont {M.}~\bibnamefont {Ejtehadi}},\ }\bibfield
  {title} {\enquote {\bibinfo {title} {Electrically rotating suspended films of
  polar liquids},}\ }\href@noop {} {\bibfield  {journal} {\bibinfo  {journal}
  {Experiments in Fluids}\ }\textbf {\bibinfo {volume} {50}},\ \bibinfo {pages}
  {419--428} (\bibinfo {year} {2011})}\BibitemShut {NoStop}%
\bibitem [{\citenamefont {Shirsavar}\ \emph {et~al.}(2012)\citenamefont
  {Shirsavar}, \citenamefont {Amjadi}, \citenamefont {Ejtehadi}, \citenamefont
  {Mozaffari},\ and\ \citenamefont {Feiz}}]{shirsavar2012}%
  \BibitemOpen
  \bibfield  {author} {\bibinfo {author} {\bibfnamefont {R.}~\bibnamefont
  {Shirsavar}}, \bibinfo {author} {\bibfnamefont {A.}~\bibnamefont {Amjadi}},
  \bibinfo {author} {\bibfnamefont {M.}~\bibnamefont {Ejtehadi}}, \bibinfo
  {author} {\bibfnamefont {M.}~\bibnamefont {Mozaffari}}, \ and\ \bibinfo
  {author} {\bibfnamefont {M.}~\bibnamefont {Feiz}},\ }\bibfield  {title}
  {\enquote {\bibinfo {title} {Rotational regimes of freely suspended liquid
  crystal films under electric current in presence of an external electric
  field},}\ }\href@noop {} {\bibfield  {journal} {\bibinfo  {journal}
  {Microfluidics and nanofluidics}\ ,\ \bibinfo {pages} {1--7}} (\bibinfo
  {year} {2012})}\BibitemShut {NoStop}%
\bibitem [{\citenamefont {Liu}\ \emph {et~al.}(2012)\citenamefont {Liu},
  \citenamefont {Zhang}, \citenamefont {Li},\ and\ \citenamefont
  {Jiang}}]{liu2012}%
  \BibitemOpen
  \bibfield  {author} {\bibinfo {author} {\bibfnamefont {Z.-Q.}\ \bibnamefont
  {Liu}}, \bibinfo {author} {\bibfnamefont {G.-C.}\ \bibnamefont {Zhang}},
  \bibinfo {author} {\bibfnamefont {Y.-J.}\ \bibnamefont {Li}}, \ and\ \bibinfo
  {author} {\bibfnamefont {S.-R.}\ \bibnamefont {Jiang}},\ }\bibfield  {title}
  {\enquote {\bibinfo {title} {Water film motor driven by alternating electric
  fields: Its dynamical characteristics},}\ }\href@noop {} {\bibfield
  {journal} {\bibinfo  {journal} {Physical Review E}\ }\textbf {\bibinfo
  {volume} {85}},\ \bibinfo {pages} {036314} (\bibinfo {year}
  {2012})}\BibitemShut {NoStop}%
\bibitem [{\citenamefont {Liu}\ \emph {et~al.}(2013)\citenamefont {Liu},
  \citenamefont {Li}, \citenamefont {Gan}, \citenamefont {Jiang},\ and\
  \citenamefont {Zhang}}]{liu2013}%
  \BibitemOpen
  \bibfield  {author} {\bibinfo {author} {\bibfnamefont {Z.-Q.}\ \bibnamefont
  {Liu}}, \bibinfo {author} {\bibfnamefont {Y.-J.}\ \bibnamefont {Li}},
  \bibinfo {author} {\bibfnamefont {K.-Y.}\ \bibnamefont {Gan}}, \bibinfo
  {author} {\bibfnamefont {S.-R.}\ \bibnamefont {Jiang}}, \ and\ \bibinfo
  {author} {\bibfnamefont {G.-C.}\ \bibnamefont {Zhang}},\ }\bibfield  {title}
  {\enquote {\bibinfo {title} {Water film washers and mixers: their rotational
  modes and electro-hydrodynamical flows induced by square-wave electric
  fields},}\ }\href@noop {} {\bibfield  {journal} {\bibinfo  {journal}
  {Microfluidics and Nanofluidics}\ ,\ \bibinfo {pages} {1--10}} (\bibinfo
  {year} {2013})}\BibitemShut {NoStop}%
\bibitem [{\citenamefont {Zahn}(2001)}]{zahn2001}%
  \BibitemOpen
  \bibfield  {author} {\bibinfo {author} {\bibfnamefont {M.}~\bibnamefont
  {Zahn}},\ }\bibfield  {title} {\enquote {\bibinfo {title} {Magnetic fluid and
  nanoparticle applications to nanotechnology},}\ }\href@noop {} {\bibfield
  {journal} {\bibinfo  {journal} {Journal of Nanoparticle Research}\ }\textbf
  {\bibinfo {volume} {3}},\ \bibinfo {pages} {73--78} (\bibinfo {year}
  {2001})}\BibitemShut {NoStop}%
\bibitem [{\citenamefont {Chaves}\ \emph {et~al.}(2006)\citenamefont {Chaves},
  \citenamefont {Rinaldi}, \citenamefont {Elborai}, \citenamefont {He},\ and\
  \citenamefont {Zahn}}]{chaves2006}%
  \BibitemOpen
  \bibfield  {author} {\bibinfo {author} {\bibfnamefont {A.}~\bibnamefont
  {Chaves}}, \bibinfo {author} {\bibfnamefont {C.}~\bibnamefont {Rinaldi}},
  \bibinfo {author} {\bibfnamefont {S.}~\bibnamefont {Elborai}}, \bibinfo
  {author} {\bibfnamefont {X.}~\bibnamefont {He}}, \ and\ \bibinfo {author}
  {\bibfnamefont {M.}~\bibnamefont {Zahn}},\ }\bibfield  {title} {\enquote
  {\bibinfo {title} {Bulk flow in ferrofluids in a uniform rotating magnetic
  field},}\ }\href@noop {} {\bibfield  {journal} {\bibinfo  {journal} {Physical
  review letters}\ }\textbf {\bibinfo {volume} {96}},\ \bibinfo {pages}
  {194501} (\bibinfo {year} {2006})}\BibitemShut {NoStop}%
\bibitem [{\citenamefont {Amirouche}, \citenamefont {Zhou},\ and\ \citenamefont
  {Johnson}(2009)}]{amirouche2009}%
  \BibitemOpen
  \bibfield  {author} {\bibinfo {author} {\bibfnamefont {F.}~\bibnamefont
  {Amirouche}}, \bibinfo {author} {\bibfnamefont {Y.}~\bibnamefont {Zhou}}, \
  and\ \bibinfo {author} {\bibfnamefont {T.}~\bibnamefont {Johnson}},\
  }\bibfield  {title} {\enquote {\bibinfo {title} {Current micropump
  technologies and their biomedical applications},}\ }\href@noop {} {\bibfield
  {journal} {\bibinfo  {journal} {Microsystem technologies}\ }\textbf {\bibinfo
  {volume} {15}},\ \bibinfo {pages} {647--666} (\bibinfo {year}
  {2009})}\BibitemShut {NoStop}%
\bibitem [{\citenamefont {Nisar}\ \emph {et~al.}(2008)\citenamefont {Nisar},
  \citenamefont {Afzulpurkar}, \citenamefont {Mahaisavariya},\ and\
  \citenamefont {Tuantranont}}]{nisar2008}%
  \BibitemOpen
  \bibfield  {author} {\bibinfo {author} {\bibfnamefont {A.}~\bibnamefont
  {Nisar}}, \bibinfo {author} {\bibfnamefont {N.}~\bibnamefont {Afzulpurkar}},
  \bibinfo {author} {\bibfnamefont {B.}~\bibnamefont {Mahaisavariya}}, \ and\
  \bibinfo {author} {\bibfnamefont {A.}~\bibnamefont {Tuantranont}},\
  }\bibfield  {title} {\enquote {\bibinfo {title} {Mems-based micropumps in
  drug delivery and biomedical applications},}\ }\href@noop {} {\bibfield
  {journal} {\bibinfo  {journal} {Sensors and Actuators B: Chemical}\ }\textbf
  {\bibinfo {volume} {130}},\ \bibinfo {pages} {917--942} (\bibinfo {year}
  {2008})}\BibitemShut {NoStop}%
\bibitem [{\citenamefont {Laser}\ and\ \citenamefont
  {Santiago}(2004)}]{laser2004}%
  \BibitemOpen
  \bibfield  {author} {\bibinfo {author} {\bibfnamefont {D.}~\bibnamefont
  {Laser}}\ and\ \bibinfo {author} {\bibfnamefont {J.}~\bibnamefont
  {Santiago}},\ }\bibfield  {title} {\enquote {\bibinfo {title} {A review of
  micropumps},}\ }\href@noop {} {\bibfield  {journal} {\bibinfo  {journal}
  {Journal of micromechanics and microengineering}\ }\textbf {\bibinfo {volume}
  {14}},\ \bibinfo {pages} {R35} (\bibinfo {year} {2004})}\BibitemShut
  {NoStop}%
\end{thebibliography}%

\end{document}